\documentclass[journal]{IEEEtran}

\usepackage{cite}
\usepackage{amsmath,amssymb,amsfonts}
\usepackage{algorithmic}
\usepackage{graphicx}
\usepackage{textcomp}
\usepackage{xcolor}
\def\BibTeX{{\rm B\kern-.05em{\sc i\kern-.025em b}\kern-.08em
    T\kern-.1667em\lower.7ex\hbox{E}\kern-.125emX}}

\usepackage{booktabs}
\usepackage{multirow}
\usepackage{pifont}
\usepackage{url}
\usepackage{balance}

\def\W{{\mathbf W}}
\def\x{{\mathbf x}}
\def\ii{{\hat{\imath}}}	
\def\ij{{\hat{\jmath}}} 
\def\ik{{\hat{\kappa}}}	
\def\bH{{\mathbb H}}
\def\bC{{\mathbb C}}
\def\bQ{{\mathbb Q}}
    
\begin{document}

\title{Attention-Map Augmentation for Hypercomplex Breast Cancer Classification}

\author{Eleonora Lopez, Filippo Betello, Federico Carmignani, Eleonora Grassucci, and Danilo Comminiello \thanks{Authors are with the Department of Information Engineering, Electronics and Telecommunications (DIET), Sapienza University of Rome, Italy.}%
\thanks{Corresponding author's email: eleonora.lopez@uniroma1.it.}}

\maketitle

\begin{abstract}

Breast cancer is the most widespread neoplasm among women and early detection of this disease is critical. 
Deep learning techniques have become of great interest to improve diagnostic performance. However, distinguishing between malignant and benign masses in whole mammograms poses a challenge, as they appear nearly identical to an untrained eye, and the region of interest (ROI) constitutes only a small fraction of the entire image. In this paper, we propose a framework, parameterized hypercomplex attention maps (PHAM), to overcome these problems. Specifically, we deploy an augmentation step based on computing attention maps. Then, the attention maps are used to condition the classification step by constructing a multi-dimensional input comprised of the original breast cancer image and the corresponding attention map. In this step, a parameterized hypercomplex neural network (PHNN) is employed to perform breast cancer classification. The framework offers two main advantages. First, attention maps provide critical information regarding the ROI and allow the neural model to concentrate on it. Second, the hypercomplex architecture has the ability to model local relations between input dimensions thanks to hypercomplex algebra rules, thus properly exploiting the information provided by the attention map. We demonstrate the efficacy of the proposed framework on both mammography images as well as histopathological ones. We surpass attention-based state-of-the-art networks and the real-valued counterpart of our approach. The code of our work is available at \url{https://github.com/ispamm/AttentionBCS}.



\end{abstract}



\section{Introduction}
\label{sec:intro}

\begin{figure}[t]
    \centering
    \includegraphics[width=0.95\linewidth]{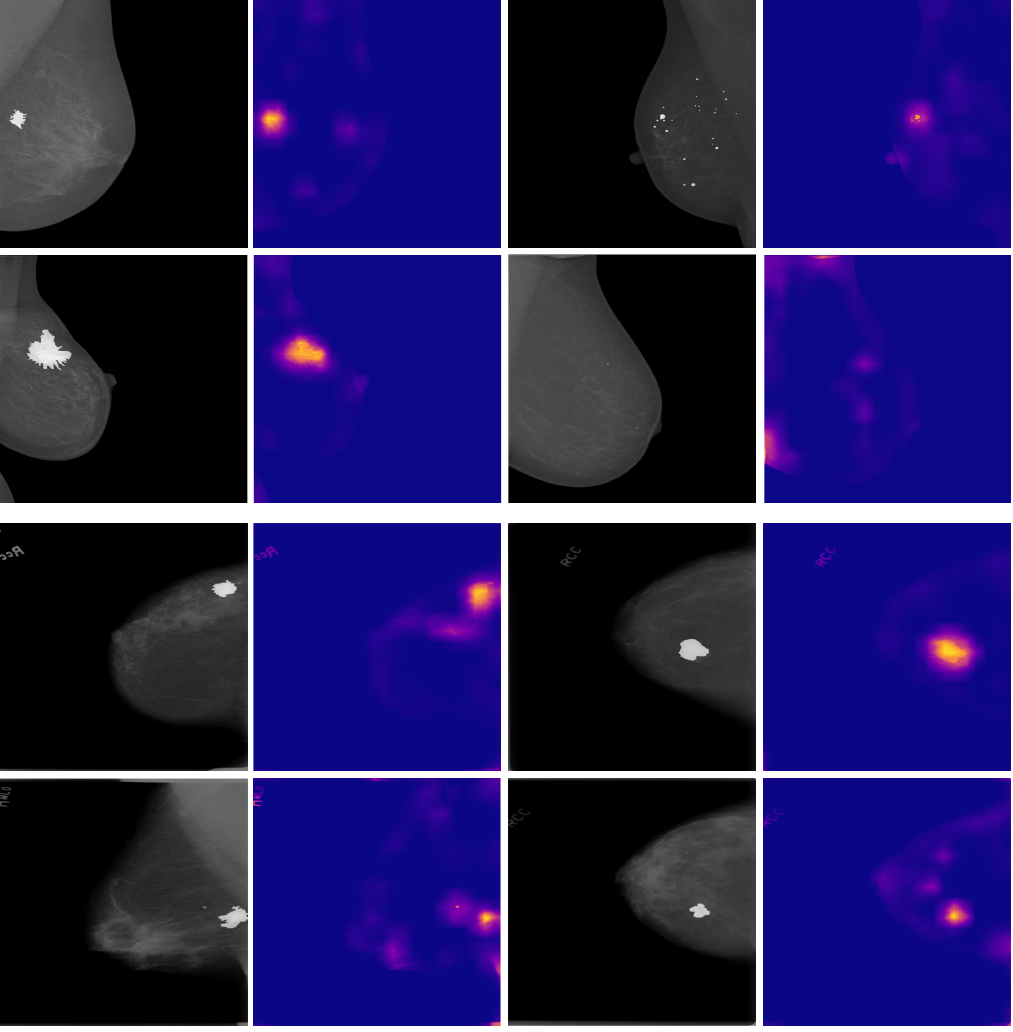}
    \caption{Top rows: attention maps of INbreast obtained from PatchConvNet fine-tuned on CBIS-DDSM. Bottom rows: attention maps of CBIS-DDSM obtained from PatchConvNet fine-tuned on INbreast. The left column comprises mammograms with a malignant finding, while the right presents negative/benign mammograms.}
    \label{fig:attn_maps}
\end{figure}

Breast cancer is the most common cancer in women worldwide. While mammography screenings have contributed to reducing mortality, there has been a gradual increase in the incidence rates of this disease since the mid-2000s \cite{siegel2023cancer}. Indeed, the mammography exam is by no means a perfect imaging test, characterized by a high rate of false positives and the subsequent need for unnecessary biopsies. Traditional computer-aided detection (CAD) algorithms fail to improve diagnostic performance and lead to high recall rates \cite{abdelhafiz2019deep}. On the other hand, deep learning-based CAD systems have been shown to succeed in assisting clinicians during the reading process, reaching a higher diagnostic accuracy \cite{eltoukhy2022classification, naguib2023classification}. This has led to an increased interest in this research area, with a variety of open problems, from reducing the number of false positives \cite{wu2021reducing} to exploiting the multi-view nature of mammography \cite{wu2020deep, lopez2022hypercomplex} and so on \cite{zhao2021diagnose}.

Nonetheless, applying deep learning techniques for this kind of problem still presents challenges. To begin with, the task itself is much more difficult if compared to the classification of natural images. Indeed, discriminating between benign and malignant tumors requires trained and expert radiologists, thus being a far from trivial problem also for neural networks \cite{murtaza2020deep}. In addition, the region of interest (ROI) comprises a tiny portion of the entire mammogram. Because of this, it is hard to detect and, even more importantly, to concentrate on it and learn to pay particular attention to such a small patch \cite{shen2019deep}. Finally, the fine-grained detail that characterizes high-resolution mammograms is crucial to identifying and correctly diagnosing masses. However, this is lost due to image resizing, a necessary step in order to train a neural model efficiently \cite{wu2020deep}.

The attention mechanism has become a successful technique to handle these kinds of challenges. It allows the network to concentrate on the most critical regions of the input \cite{vaswani2017attention}. Typically, it is found inside transformer-like architectures \cite{dalmaz2022resvit, mo2023hover}. However, new strategies are being developed to endow convolutional neural networks (CNNs) with attention layers \cite{touvron2021augmenting, woo2018cbam}.

In this paper, we introduce a novel approach to exploit the information learned by attention layers to address the aforementioned challenges related to breast cancer screening. We propose a framework that consists of an attention-map augmentation step and a parameterized hypercomplex network as backbone model for breast cancer classification. For simplicity, we refer to it as the parameterized hypercomplex attention maps (PHAM) framework. In detail, the attention-map augmentation step consists in computing attention maps for each cancer image with an already existing model, e.g., PatchConvNet \cite{touvron2021augmenting}, such as the ones displayed in Fig.~\ref{fig:attn_maps}. Then we employ them as additional input to the hypercomplex backbone. In this way, we perform a form of conditioning on the attention map during the classification step. Thus, we provide the neural network information regarding the abnormal regions of the breast cancer image, which are most significant for diagnosis. 
We exploit the new multi-dimensional input through parameterized hypercomplex networks. Indeed, these architectures have the ability to model correlations in multi-dimensional data while also obtaining a more lightweight network \cite{grassucci2021phnns, lopez2022hypercomplex}. Certainly, quaternion and generalized hypercomplex networks have gained much interest in the past few years \cite{comminiello2019quaternion, parcollet2019qcnn, zhang2021phm}. The success of quaternion neural networks (QNNs) is owed to quaternion algebra that allows to model both global and local relations in input $4$D data \cite{parcollet2019survey, parcollet2019qrnn}. This, in turn, allows to learn a more powerful representation in the latent space \cite{brignone2022efficient}. Thereafter, parameterized hypercomplex neural networks (PHNNs) were introduced in order to bring the advantages of QNNs to general input domains of any dimensionality $n$ \cite{ grassucci2021phnns, zhang2021phm}. 

Thus, in this work, we employ parameterized hypercomplex (PH) ResNets as the backbone of our framework. In this way, we are able to capture the correlations between the original breast cancer image and the corresponding attention map thanks to hypercomplex algebra properties. Through an experimental analysis conducted on public benchmark datasets of mammograms, i.e. CBIS-DDSM \cite{lee2017cbis} and INbreast \cite{morerira2012inbreast}, and histopathological microscopic images, i.e., BreakHis \cite{spanhol2016breakhis}, we show how the proposed method is able to outperform the real-valued counterpart as well as attention-based state-of-the-art models.

The rest of the paper is organized as follows. Section~\ref{sec:related} lays out an overview of the current related works. Section~\ref{sec:method} gives a detailed description of the proposed framework and the theory behind hypercomplex networks. Section~\ref{sec:exp_eval} provides technical details regarding data, training, and experimental results. Finally, conclusions are drawn in Section~\ref{sec:conclusions}.

\section{Related works}
\label{sec:related}

Prior works adopt several different approaches for the task of breast cancer classification. Owing to the aforementioned problems, many studies focus on the classification of single patches instead of the whole mammogram \cite{kooi2017classifying, kooi2017large, al2022hybrid, alhussan2023classification}. Certainly, by considering the single patch containing the ROI, there is no need to devise a method for the network to focus on it since it becomes the main object in the image. Moreover, the details would be clearly visible even after image resizing, thus making the classification task much easier compared to whole-mammogram classification \cite{shen2019deep}. On the other hand, methods that directly process the whole mammography image, usually adopt a pertaining strategy based on either patch-level classification or on natural images in order to alleviate these problems \cite{wu2021reducing, shen2019deep, wu2020deep, lopez2022hypercomplex, li2021dual, wang2023information, jabeen2023bc2netrf}. Instead, recent approaches for breast cancer classification from histopathology images involve graph neural networks \cite{patel2023garl}, architectures specifically tailored for histopathology \cite{eltoukhy2022classification} and transfer learning \cite{hamedani2023breast}.

More recently, with the success of the transformer \cite{vaswani2017attention} and vision transformer (ViT) \cite{dosovitskiy2020vit} architectures, also the medical imaging community has started to develop architectures based on the former models \cite{dalmaz2022resvit, mo2023hover, zhou2023nnformer}. Indeed, these methods aim to focus the attention of the neural model on the ROI, which is often small in the medical scenario, through the self-attention mechanism. 
However, transformer-based models are characterized by much less inductive biases which instead are inherent to convolutions, i.e., locality and translation equivariance \cite{chen2021when, peng2021conformer}. For this reason, many works started to investigate strategies to incorporate self-attention into convolutional neural networks (CNNs). The goal is to maintain its intrinsic inductive biases and additionally gain the advantages of the attention mechanism. One of the first popular works introduced a convolutional bottleneck attention module (CBAM), which can be easily plugged into any CNN to refine feature maps through the inferred channel and spatial attention maps \cite{woo2018cbam}. Then, a more recent study proposed an extension of this module specifically designed for breast cancer classification that is able to exploit cross-view information \cite{zhao2020crossview}. The same strategy is also being employed for various medical areas such as EEG decoding \cite{ma2023temporal} and Alzheimer's disease classification \cite{hu2023conv}. Finally, Touvron et al. \cite{touvron2021augmenting} introduced an attention-based aggregation layer to augment any CNN. Furthermore, they proposed a patch-based architecture, PatchConvNet, that allows to obtain higher quality attention maps by keeping the input resolution constant throughout the network. In this paper, we take a different approach and employ the pretrained PatchConvNet to obtain attention maps for mammography images. Then, we exploit them through parameterized hypercomplex layers in order to overcome the problems presented in Section~\ref{sec:intro}.

\section{Proposed method}
\label{sec:method}

\begin{figure*}[t]
    \centering
    \includegraphics[width=\textwidth]{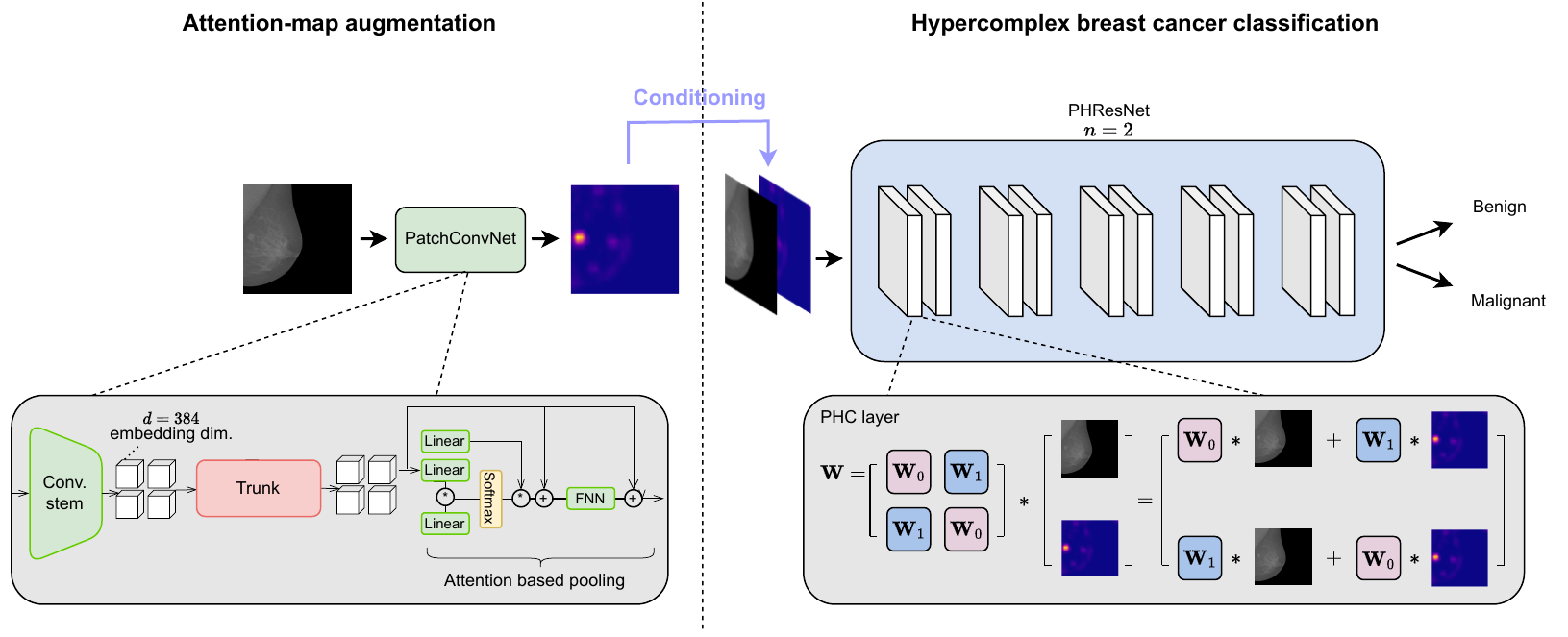}
    \caption{PHAM framework. On the left, the attention-map augmentation step is depicted. Herein, attention maps are computed offline with the fine-tuned PatchConvNet model. Then, they are used to perform a form of conditioning on the hypercomplex model. On the right, a PHResNet with $n=2$ for mammography images ($n=4$ for histopathology images) is employed as the backbone to perform breast cancer classification. It is able to model relations between breast cancer images and attention maps, as can be seen in the visualization of the PHC layer.}
    \label{fig:method}
\end{figure*}

In this section, we expound on the proposed method, parameterized hypercomplex attention maps (PHAM) framework for breast cancer detection, depicted in Fig.~\ref{fig:method}. Initially, we present the theoretical background of hypercomplex neural networks, followed by a detailed description of the framework we introduce.

\subsection{Parameterized hypercomplex models}

Quaternion neural networks (QNNs) are models that operate in an extension of complex numbers $\bC$, namely the quaternion domain $\bQ$. A quaternion is defined as $q = q_0 + q_1 \ii + q_2 \ij + q_3 \ik$, in which $q_i \in \mathbb R$, with $i=0,\ldots,3$, are the real coefficients and $\ii, \ij, \ik \in \bQ$ are the imaginary units. The product between two imaginary units is not commutative, thus the Hamilton product has been introduced to properly model the multiplication of two quaternions. Thanks to this product, the weight matrix and the input can be encapsulated into a quaternion as 

\begin{equation}
\begin{split}
    & \W = \W_0 + \W_1 \ii + \W_2 \ij + \W_3 \ik \\
    & \x = \x_0 + \x_1\ii + \x_2\ij + \x_3\ik.
\end{split}
\end{equation}

\noindent Then the convolution between them is defined as follows: 

\begin{equation}
{\bf{W}} * {\bf{x}} = \left[ {\begin{array}{*{20}c}
   \hfill {{\bf{W}}_0 } & \hfill { - {\bf{W}}_1 } & \hfill { - {\bf{W}}_2 } & \hfill { - {\bf{W}}_3 } \\
   \hfill {{\bf{W}}_1 } & \hfill {{\bf{W}}_0 } & \hfill { - {\bf{W}}_3 } & \hfill {{\bf{W}}_2 } \\
   \hfill {{\bf{W}}_2 } & \hfill {{\bf{W}}_3 } & \hfill {{\bf{W}}_0 } & \hfill { - {\bf{W}}_1 } \\
   \hfill {{\bf{W}}_3 } & \hfill { - {\bf{W}}_2 } & \hfill {{\bf{W}}_1 } & \hfill {{\bf{W}}_0 } \\
\end{array}} \right] * \left[ {\begin{array}{*{20}c}
   {{\bf{x}}_0 } \hfill  \\
   {{\bf{x}}_1 } \hfill  \\
   {{\bf{x}}_2 } \hfill  \\
   {{\bf{x}}_3 } \hfill  \\
\end{array}} \right].
\label{eq:qconv}
\end{equation}

\noindent In this way, the filter submatrices $\W_i, \; i=0,\ldots,3$ in eq.~\eqref{eq:qconv} are shared between each dimension of the input $\x$. This property brings two main advantages. Firstly, the number of parameters is reduced by $1/4$, thus yielding a much lighter model with respect to the real-valued counterpart. Secondly, by sharing weights between input dimensions, the neural model is endowed with the ability to model local relations. Therefore, QNNs, in addition to modeling global relations as standard neural networks are able to do, also capture correlations among channels by treating the input as a unique entity. Instead, real-valued networks assign different weights to different dimensions, thus treating them independently when they are actually correlated. Indeed, this additional information, which networks in the real domain fail to grasp, allows quaternion models to learn a more powerful representation of the data and yield more accurate predictions as a result.

However, in order to extend this approach to inputs of any dimensionality $n$D, instead of being limited to $4$D inputs, a generalization in the hypercomplex domain $\bH$ was introduced, i.e., parameterized hypercomplex neural networks (PHNNs) \cite{zhang2021phm, grassucci2021phnns}. In this case, the filter weight matrix $\W$ is expressed as a parameterized sum of Kronecker products \cite{grassucci2021phnns}:

\begin{equation}
    \W = \sum_{i=0}^n \mathbf{A}_i \otimes \mathbf{F}_i,
\label{eq:phc}
\end{equation}

\noindent where $n$ defines the number domain in which the network operates (e.g., $n=4$ corresponds to the quaternion domain) and can be chosen freely in order to best represent the input data. While matrices $\mathbf{A}_i$ and $\mathbf{F}_i$ encode the algebra rules and the weight filters, respectively, both learned during training. With this formulation, the advantages of quaternion models are maintained, thus still leveraging hypercomplex algebra properties to model latent relations among channels, while reducing the number of free parameters by $1/n$.

\subsection{Parameterized hypercomplex attention maps (PHAM)}

We design the parameterized hypercomplex attention maps (PHAM) framework as follows. We define an augmentation operation based on the computation of attention maps for each input image. Then we use them to condition the hypercomplex model during training to improve the classification performance of breast cancer. The relations among the original images and the relative attention maps are exploited through parameterized hypercomplex models thanks to their aforementioned properties.

Indeed, any neural method can be easily defined in the hypercomplex domain \cite{grassucci2021phnns}, therefore in this paper, we deploy parameterized hypercomplex ResNets (PHResNets). In such a manner, the standard residual block $\mathbf{y} = \mathcal{F}(\mathbf{x}) + \mathbf{x}$, where $\mathcal{F}$ is composed of interleaving convolutional layers, batch normalization (BN) and ReLu activation function, becomes:
\begin{equation}
    \mathcal{F}(\mathbf{x}) = \text{BN} \left(\text{PHC} \left(\text{ReLU} \left(\text{BN} \left(\text{PHC} \left(\mathbf{x} \right)\right)\right)\right)\right).
\label{eq:phresnet}
\end{equation}

\noindent Herein, PHC refers to parameterized hypercomplex convolutions, with the weight matrix defined as in Eq.~\eqref{eq:phc}, and $\mathbf{x}$ is the multidimensional input composed of breast cancer images and the corresponding attention map. 

Attention maps are a visualization of what the attention layer has learned during training \cite{touvron2021augmenting}. We propose to exploit this information in image form, i.e., we augment the dataset with the attention maps, inspired by the similar utilization of heatmaps in recent works \cite{wu2020deep, singh2023flowgrad}. More in detail, to compute the attention maps we deploy the recent PatchConvNet \cite{touvron2021augmenting}, since it allows to obtain high-resolution attention maps thanks to its non-hierarchical design. 
As the network is initially trained on ImageNet, direct utilization for medical images is not feasible. Thus, we first perform a fine-tuning step on a breast cancer dataset and thereafter we apply the fine-tuned model to infer attention maps on other breast cancer databases. Then, we construct the augmented dataset in which each sample is composed of the original mammograms or histopathological images and the corresponding attention maps, considering them as a single multi-dimensional input. In this way, the attention map conditions the training process by emphasizing the most critical portion of the image. This allows the neural network to focus on these crucial areas, thereby enhancing its predictive capabilities and leading to improved performance.

To conclude, when processing images corresponding to mammograms, the input $\mathbf{x}$ in Eq.~\eqref{eq:phresnet} has two dimensions. The first represents the mammogram and the second corresponds to the attention map. We set the hyperparameter $n=2$ and operate in the complex domain $\bC$ in order to capture relations between them. On the other hand, when considering histopathological images, we set $n=4$, thus operating in the quaternion domain. This choice aligns with the fact that histology images are stored in RGB format, comprising $3$ channels, plus the additional attention map. According to the aforementioned discussions, by endowing the architecture with PHC layers, we can better leverage the attention maps. Indeed, thanks to hypercomplex algebra properties, the hypercomplex network has the ability to capture local relations between original breast cancer images and the respective maps. In this way, we truly exploit the additional information contained in the attention maps, unlike real-valued networks which fail to model local relations \cite{parcollet2019survey}.

\section{Experimental evaluation}
\label{sec:exp_eval}

\subsection{Data}

We validate our proposed approach on publicly available datasets of mammography, CBIS-DDSM \cite{lee2017cbis} and INbreast \cite{morerira2012inbreast}, and histopathology, BreakHis \cite{spanhol2016breakhis},  widely used in literature \cite{eltoukhy2022classification, shen2019deep}.

The Curated Breast Imaging Subset of the Digital Database for Screening Mammography (CBIS-DDSM) \cite{lee2017cbis} consists of scanned film mammography images standardized in the DICOM format. It provides biopsy-proven pathology labels, which can be either benign or malignant. We utilize the official training and testing data splits and obtain the validation split by performing an additional stratified partition of the training set, for a total of $991$ images for training, $240$ for validation and $361$ for testing. 

On the other hand, INbreast \cite{morerira2012inbreast} is a much smaller database consisting of $410$ full-filed digital mammography images. Pathological labels are not available but instead, it provides BI-RADS classification from which we extract binary labels by considering categories $1$ and $2$ as negative, and $4$, $5$ and $6$ as positive, discarding category $3$ \cite{shen2019deep}. We split the dataset in a stratified fashion and take $20\%$ of images for validation and testing, respectively.

The third dataset, the Breast Cancer Histopathological Image Classification (BreakHis) \cite{spanhol2016breakhis}, consists of $9,109$ microscopic images of breast tissue at four different magnifying factors, i.e., $40$X, $100$X, $200$X and $400$X. BreakHis is divided into two main groups, that is benign and malignant tumors. We create different sets for training, validation and testing by splitting the dataset patient-wise in order to avoid information-leaking, taking $20\%$ for the two latter sets. 

The same preprocessing is applied for all datasets: images are resized to $384 \times 384$ and standardized. Data augmentation operations are applied for training images only, i.e., random horizontal and vertical flips and a random rotation of degrees taken from $(-10^\circ, +10^\circ)$.

\subsection{Validation metrics}

To validate the experiments conducted on datasets of mammogram images we utilize the area under the ROC curve (AUC), as it is the primary metric employed in the literature \cite{wu2020deep, wu2021reducing, shen2019deep}. It provides a measure of the predictive ability of a classifier at different probability thresholds, taking into account the trade-off between true positive and false positive rates. For the second set of experiments, the metric adopted in the original paper is the accuracy \cite{spanhol2016breakhis}, thus we also utilize it for evaluating the performance of the proposed approach on the BreakHis dataset.

\subsection{Training and architecture details}

ResNet-based architectures are trained with the Adam optimizer \cite{Diederik2015adam-opt},  which is a variant of Stochastic Gradient Descent \cite{ruder2016overview} with adaptive moment estimation, a learning rate of $10^{-5}$ and weight decay at $5 \times 10^{-4}$. The number of epochs is set to $100$ and we early stop the training when the AUC on the validation data does not improve for $20$ epochs. Instead, PatchConvNet is fine-tuned with the Lamb optimizer \cite{you2020lamb}, a variant of Adam that provides a strategy for adapting the learning rate in large batch settings. We set the learning rate at $5 \times 10^{-4}$ and weight decay at $10^{-2}$, following the recipe for fine-tuning experiments of the original paper. Moreover, we employ the s$60$ configuration of PatchConvNet, which consists of an embedding dimension of $384$ and $60$ repeated blocks in the trunk \cite{touvron2021augmenting}. Finally, for hypercomplex models, we employ ResNet$18$ and ResNet$50$ in the hypercomplex domain for the first and second sets of experiments, respectively, as BreakHis is a much larger dataset compared to INbreast and CBIS-DDSM.

\subsection{Experiments and results}
\begin{table*}[t]
\centering
\caption{Results on the test sets of the INbreast and CBIS-DDSM datasets. Attention maps (AM) are obtained from PatchConvNet fine-tuned on CBIS-DDSM (top) and INbreast (bottom), respectively. Results in bold and underlined correspond to the best and second best.}
\label{tab:inbreast}
\begin{tabular}{llcccc}
\toprule
\multicolumn{1}{l}{Dataset} & \multicolumn{1}{l}{Model} & \multicolumn{1}{c}{Params} & \multicolumn{1}{c}{PH} & \multicolumn{1}{c}{AM} & \multicolumn{1}{c}{AUC} \\  \midrule 
\multirow{6}{*}{INbreast} & ResNet18 & 11M & \ding{55} & \ding{55} & 0.480 $\pm$ 0.070 \\
& ResNet18+CBAM \cite{zhao2020crossview} & 11M & \ding{55} & \ding{55} & 0.491 $\pm$ 0.135 \\
& PatchConvNet \cite{touvron2021augmenting} & 24M & \ding{55} & \ding{55} & 0.806 $\pm$ 0.218 \\
& ResNet18 & 11M & \ding{55} & \ding{51} & 0.840 $\pm$ 0.027 \\ 
& ResNet18+CBAM \cite{woo2018cbam} & 11M & \ding{55} & \ding{51} & \underline{0.847} $\pm$ 0.018 \\
& PHResNet18 ($n=2$) & 5M & \ding{51} & \ding{51} & \textbf{0.852} $\pm$ 0.037 \\ 
\midrule
\multirow{3}{*}{CBIS-DDSM} & ResNet18 & 11M & \ding{55} & \ding{55} & 0.659 $\pm$ 0.008 \\
& ResNet18 & 11M & \ding{55} & \ding{51} & \underline{0.694} $\pm$ 0.034 \\ 
& PHResNet18 ($n=2$) & 5M & \ding{51} & \ding{51} & \textbf{0.725} $\pm$ 0.027 \\
\bottomrule   
\end{tabular}
\end{table*}

\begin{figure}[t]
    \centering
    \includegraphics[width=0.99\linewidth]{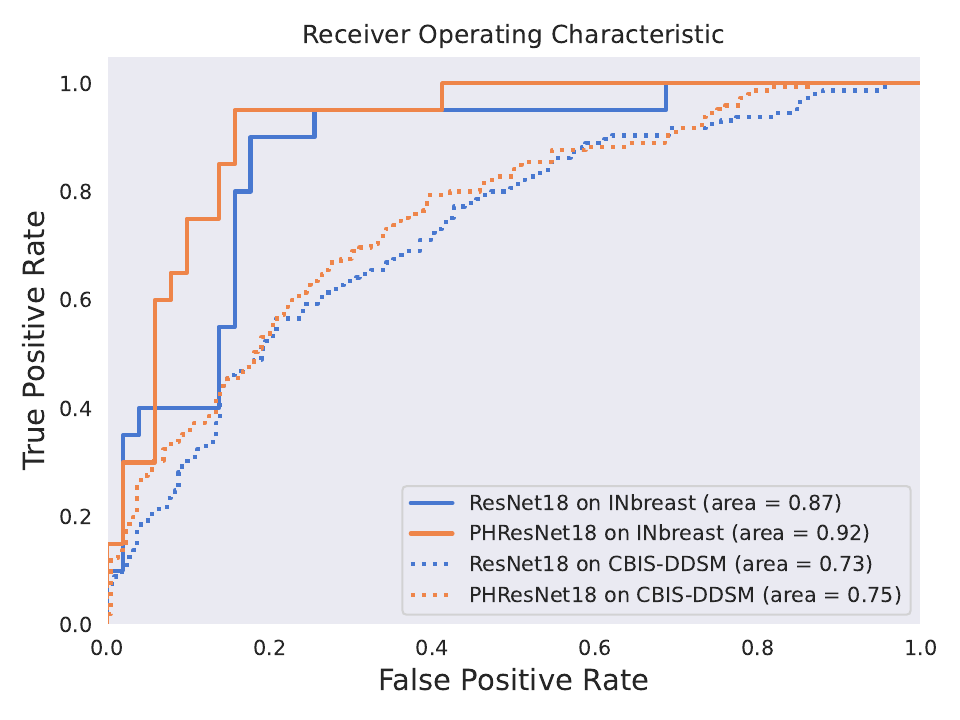}
    \caption{ROC curve corresponding to the best run of models with AM for experiments conducted on INbreast and CBIS-DDSM datasets.}
    \label{fig:roc}
\end{figure}

\begin{figure}[t]
    \centering
    \includegraphics[width=0.99\linewidth]{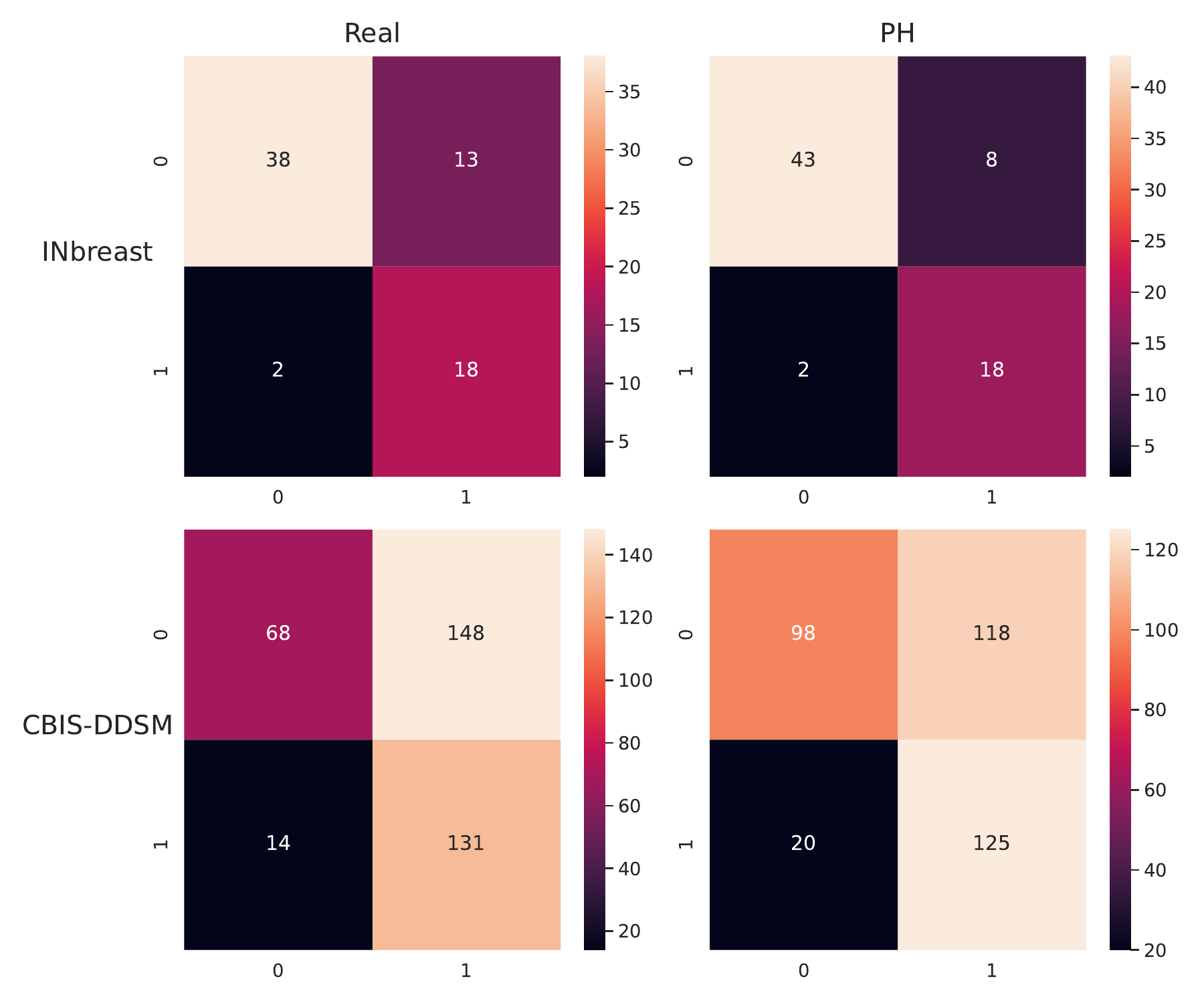}
    \caption{Confusion matrices of models with AM for experiments on INbreast and CBIS-DDSM.}
    \label{fig:cm_mamm}
\end{figure}

The first set of experiments is conducted on datasets of mammography exams. Specifically, in the first step, we fine-tune the PatchConvNet architecture on CBIS-DDSM utilizing the available ImageNet weights \cite{touvron2021augmenting}. As a second step, we operate the model at inference time to obtain the attention maps on INbreast. Finally, we train the hypercomplex network, i.e. PHResNet18 ($n=2$) with the conditioning provided by the attention map, which from here on we denote with AM. We compare the results of our approach against a baseline ResNet18 without AM, state-of-the-art methods, and the real-valued counterpart of the proposed framework, i.e., ResNet18 with AM. Notably, we do not need to test PHResNet18 without AM, as in that case the parameter $n$ would be set to $1$, which is equivalent to its real-valued counterpart, i.e. ResNet18 \cite{grassucci2021phnns}. The first state-of-the-art method we consider is the cross-view attention module (CvAM) designed for multi-view analysis of mammography, which is integrated inside a ResNet architecture \cite{zhao2020crossview}. In order to compare it with our method, we utilize the single-view equivalent of this approach, which is the CBAM module integrated in the same fashion as described in the original paper \cite{zhao2020crossview}. The second state-of-the-art network for comparison is PatchConvNet itself, directly fine-tuned on INbreast. The mean AUC over $5$ runs is reported in Tab.~\ref{tab:inbreast} together with the number of parameters of each network. Evidently, our approach in both hypercomplex and real domains outperforms both baseline and state-of-the-art methods. This shows that the proposed strategy for exploiting such knowledge is effective, thus resulting in more performant classifiers. Moreover, the proposed framework produces the most discriminant model, i.e. PHResNet with AM, which yields an AUC of $0.852$. It is also important to note that it achieves such result with just $5$M parameters, that is $1/5$ of PatchConvNet. This is thanks to hypercomplex algebra rules which allow modeling local relations between input dimensions, thus grasping correspondences among mammogram and attention map. To conclude, even though the main gain is attained by including attention maps, by introducing hypercomplex algebra the performance is still slightly improved but with a much lighter model. 

\begin{table*}[t]
\centering
\caption{Results on the test set of the BreakHis dataset at magnification factors 100X, 200X, and 400X. Attention maps (AM) are obtained from PatchConvNet fine-tuned at magnification factor 40X. Results in bold and underlined correspond to the best and second best, respectively.}
\label{tab:breakhis}
\begin{tabular}{clcccc}
\toprule
\multicolumn{1}{c}{Magnifying factor} & \multicolumn{1}{l}{Model} & \multicolumn{1}{c}{Params} & \multicolumn{1}{c}{PH} & \multicolumn{1}{c}{AM} & \multicolumn{1}{c}{Accuracy (\%)} \\  
\midrule 
\multirow{4}{*}{100X} & ResNet50 & 16M & \ding{55} & \ding{55} & 0.696 $\pm$ 0.108 \\
& PHResNet50 ($n=3$) & 5M & \ding{51} & \ding{55} & 0.767 $\pm$ 0.074 \\
& ResNet50 & 16M & \ding{55} & \ding{51} & \underline{0.809} $\pm$ 0.016 \\ 
& PHResNet50 ($n=4$) & 4M & \ding{51} & \ding{51} & \textbf{0.821} $\pm$ 0.027 \\ 
\midrule
\multirow{4}{*}{200X} & ResNet50 & 16M & \ding{55} & \ding{55} & 0.743 $\pm$ 0.020 \\
& PHResNet50 ($n=3$) & 5M & \ding{51} & \ding{55} & 0.753 $\pm$ 0.057 \\
& ResNet50 & 16M & \ding{55} & \ding{51} & \underline{0.766} $\pm$ 0.007 \\ 
& PHResNet50 ($n=4$) & 4M & \ding{51} & \ding{51} & \textbf{0.781} $\pm$ 0.016 \\ 
\midrule
\multirow{4}{*}{400X} & ResNet50 & 16M & \ding{55} & \ding{55} & 0.714 $\pm$ 0.043 \\
& PHResNet50 ($n=3$) & 5M & \ding{51} & \ding{55} & 0.730 $\pm$ 0.027 \\
& ResNet50 & 16M & \ding{55} & \ding{51} & \underline{0.746} $\pm$ 0.018 \\ 
& PHResNet50 ($n=4$) & 4M & \ding{51} & \ding{51} &\textbf{0.765} $\pm$ 0.008 \\ 
\bottomrule   
\end{tabular}
\end{table*}

Thereafter, we also perform the same experimental evaluation as described above but with the two datasets switched. Thus, PatchConvNet is fine-tuned on INbreast, then the attention maps for CBIS-DDSM are inferred and used to train the different networks. The results reported in the bottom part of Tab.~\ref{tab:inbreast} support our theory, this time even attaining much more gain from the introduction of hypercomplex algebra, i.e., from 0.694 to 0.725. Furthermore, the ROC curve for experiments in both datasets is depicted in Fig.~\ref{fig:roc} and is in accordance with the values presented in Tab.~\ref{tab:inbreast}.

\begin{figure}[t]
    \centering
    \includegraphics[width=0.99\linewidth]{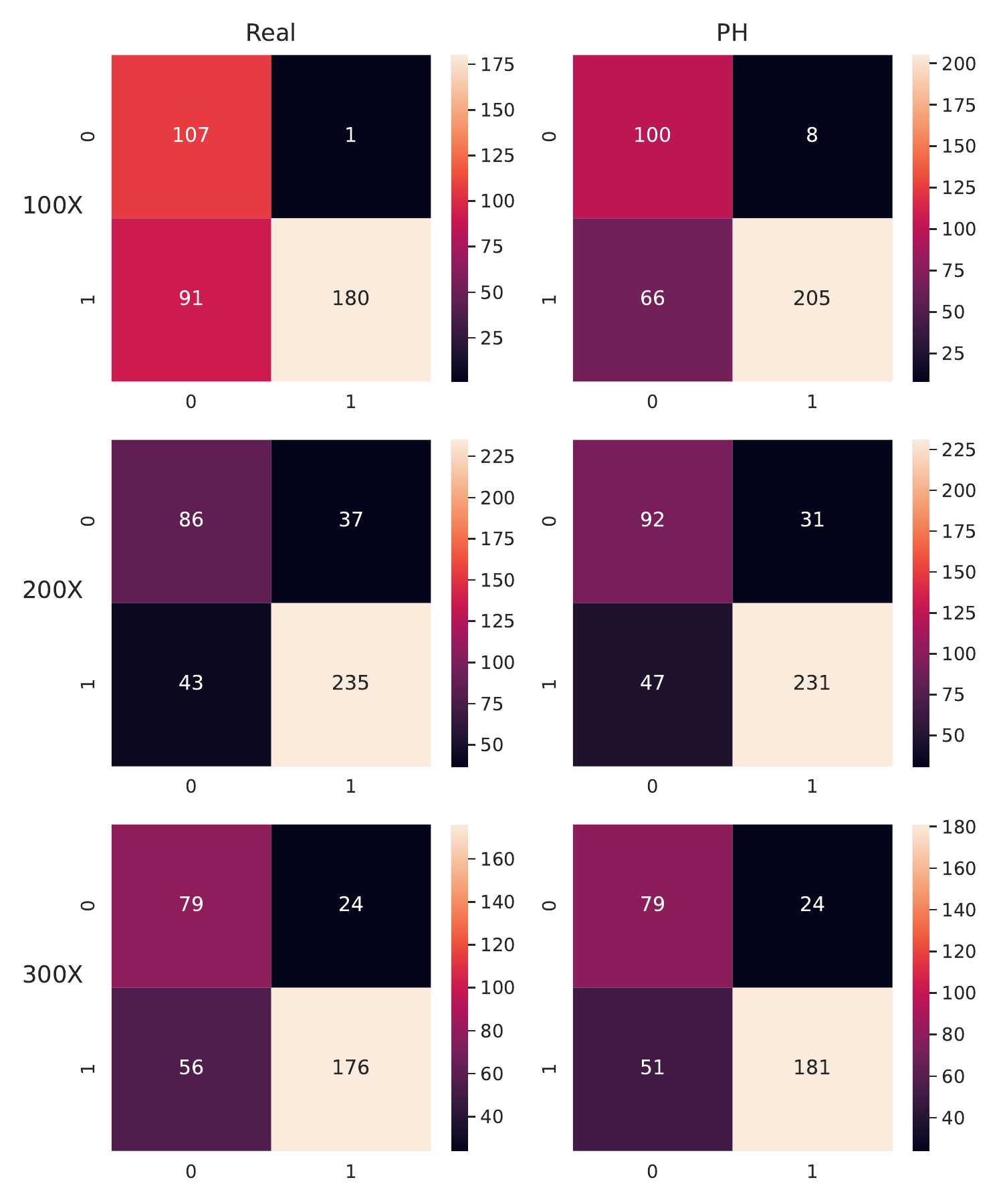}
    \caption{Confusion matrices of models with AM for experiments on BreakHis.}
    \label{fig:cm_hist}
\end{figure}

The second set of experiments is conducted on histopathological microscopic images of tumor tissue at different magnifying factors. In detail, we fine-tune PatchConvNet on images at a magnification factor of $40$X and then we utilize it at inference time to compute attention maps for the remaining magnifying factors, i.e. $100$X, $200$X, and $400$X. Finally, we train PHResNet50 with AM on the latter datasets and compare it against a vanilla ResNet50, PHResNet50 with $n=3$ (since they are RGB images, i.e. with three channels), and the real-valued respective of our method, i.e., ResNet50 with AM. The results are reported in Tab.~\ref{tab:breakhis}, showing the average of the accuracy over $3$ runs and the standard deviation. Firstly, as expected, the advantage brought by hypercomplex algebra can be seen in both scenarios with and without AM. In both cases, the mean accuracy is improved and the models are comprised of only 5M and 4M parameters, for $n=3$ and $n=4$, respectively. Secondly, the experiments demonstrate how attention maps are generalized for different magnifying factors, thus improving the performance in every case. The hypercomplex network yields the best accuracy for each scenario with a quarter of the parameters with respect to its counterpart in the real domain. Thus, we further demonstrate the efficacy of the proposed framework on microscopic images at different magnification factors of tumor tissue in addition to X-ray mammogram exams. Finally, for both sets of experiments, we illustrate the respective confusion matrices in Fig.~\ref{fig:cm_mamm} and Fig.~\ref{fig:cm_hist}. We can see that in every case either the number of false negatives or false positives is reduced by the PH model, aligning with the results of Tab.~\ref{tab:inbreast} and Tab.~\ref{tab:breakhis}.

To conclude, Table~\ref{tab:inbreast} also includes different ablation experiments as the gain from each proposed component, i.e. conditioning on attention maps (AM) and hypercomplex algebra (PH) is shown. In fact, we test a baseline ResNet, then ResNet with AM and finally, we add hypercomplex algebra with PHResNet with AM. Table~\ref{tab:breakhis} also presents the experiment with hypercomplex algebra and without AM, i.e. PHResNet with $n=3$.

\section{Conclusions}
\label{sec:conclusions}

In this paper, we have proposed a novel framework to exploit the information learned by the attention mechanism. We build an augmented dataset where each sample comprises the original image and the respective attention map. This new multi-dimensional input is used to condition the training of a PHResNet which handles it as a single unit. Thanks to hypercomplex algebra properties, the PH model has the capacity to capture latent relations between the original breast cancer image and the attention map. In this way, we effectively exploit the additional information provided by conditioning on the attention map regarding the location of the tumor region, shifting the focus of the network on it. We demonstrate the validity of the proposed framework on breast cancer datasets comprising mammography exams and histopathological microscopic images. Our approach outperforms attention-based state-of-the-art architectures and the real-valued counterpart of the proposed technique. In future works, we aim to overcome the limitations of the proposed framework which requires a fine-tuning step of a network such as PatchConvNet due to the domain difference between natural and medical images.

\section*{Acknowledgments}
This work was partially supported by the Italian Ministry of University and Research (MUR) within the PRIN 2022 Program for the project ``EXEGETE: Explainable Generative Deep Learning Methods for Medical Signal and Image Processing", under grant number 2022ENK9LS, CUP C53D23003650001, and in part by the European Union under the National Plan for Complementary Investments to the Italian National Recovery and Resilience Plan (NRRP) of NextGenerationEU,  Project PNC 0000001 D3 4 Health - SPOKE 1 - CUP B53C22006120001.

\balance
\bibliographystyle{elsarticle-num}
\bibliography{ABCS.bib}

\end{document}